\documentclass[
 reprint,
superscriptaddress,
  aps,
superscriptaddress,
 amsmath,amssymb,
 longbibliography
]{revtex4-1}

\usepackage{graphicx}
\usepackage{dcolumn}
\usepackage{bm}
\usepackage{amsmath,amssymb,amstext,amsthm,mathtools}
\usepackage{braket}
\usepackage{bbold}
\usepackage{bbm}
\usepackage{xcolor}
\usepackage{ gensymb }
\usepackage{soul}
\usepackage{subcaption}
\usepackage{mathrsfs}
\usepackage{float}
\usepackage{placeins}

\begin{document}

\setlength{\abovedisplayskip}{1pt}

\title{Generalizing the Quantum Information Model for Dynamic Diffraction}
\author{O. Nahman-Lévesque} 
\affiliation{Institute for Quantum Computing, University of Waterloo,  Waterloo, ON, Canada, N2L3G1}
\affiliation{Department of Physics, University of Waterloo, Waterloo, ON, Canada, N2L3G1}

\author{D. Sarenac}
\affiliation{Institute for Quantum Computing, University of Waterloo,  Waterloo, ON, Canada, N2L3G1}

\author{D. G. Cory}
\affiliation{Institute for Quantum Computing, University of Waterloo,  Waterloo, ON, Canada, N2L3G1}
\affiliation{Department of Chemistry, University of Waterloo, Waterloo, ON, Canada, N2L3G1}

\author{B. Heacock}
\affiliation{National Institute of Standards and Technology, Gaithersburg, Maryland 20899, USA}

\author{M. G. Huber}
\affiliation{National Institute of Standards and Technology, Gaithersburg, Maryland 20899, USA}

\author{D. A. Pushin}
\email{dmitry.pushin@uwaterloo.ca}
\affiliation{Institute for Quantum Computing, University of Waterloo,  Waterloo, ON, Canada, N2L3G1}
\affiliation{Department of Physics, University of Waterloo, Waterloo, ON, Canada, N2L3G1}

\date{\today}

\begin{abstract}

The development of novel neutron optics devices that rely on perfect crystals and nano-scale features are ushering a new generation of neutron science experiments, from fundamental physics to material characterization of emerging quantum materials. However, the standard theory of dynamical diffraction (DD) that analyzes neutron propagation through perfect crystals does not consider complex geometries, deformations, and/or imperfections which are now becoming a relevant systematic effect in high precision interferometric experiments. In this work, we expand upon a quantum information (QI) model of DD that is based on propagating a particle through a lattice of unitary quantum gates. We show that the model output is mathematically equivalent to the spherical wave solution of the Takagi-Taupin equations when in the appropriate limit, and that the model can be extended to the Bragg as well as the Laue-Bragg geometry where it is consistent with experimental data. The presented results demonstrate the universality of the QI model and its potential for modeling scenarios that are beyond the scope of the standard theory of DD.

\end{abstract}

\pacs{Valid PACS appear here}

\maketitle

\section{Introduction}

Many thermal and cold neutron instruments and experimental methods rely on Bragg diffraction from nearly-perfect crystals.  These include neutron interferometers \cite{rauch1974test, Klepp_2014, rauch2015neutron, pushin2015neutron, huber2019overview}, Bonse-Hart double crystal diffractometers \cite{bonse1965tailless, barker2005design}, storage cavities \cite{schuster1990test}, spin-rotating channels \cite{gentile2019study}, spin rotation in non-centrosymmetric crystals \cite{fedorov2010measurement}, and high-precision structure factor measurements \cite{Shull_perfect_crystals, heacock2021pendell}. The theory of dynamical diffraction, which was originally  developed by Cowley and Moodie for electron propagation through lattices \cite{cowley1957scattering}, describes the behavior of neutrons inside perfect crystals and must be used over the kinematic theory when the crystal thickness or mosaic block size is larger than the extinction length \cite{sears1978dynamical, abov2002dynamical,lemmel2013influence,lemmel2007dynamical}.
However, use of the standard theory can only reasonably accommodate relatively-simple crystal geometries \cite{mocella2008experimental}, strain fields \cite{Takagi:a03704,voronin2017effect}, and incoming beam phase spaces \cite{klein1983longitudinal, pushin_prl_coherence:250404}, factors which impact device design and can bias experimental results \cite{Shull_perfect_crystals, barker2005design, heacock2017neutron, gentile2019study, saggu2016decoupling}. 

Nsofini et al. \cite{Nsofini_2016,nsofini2017noise,nsofini2019coherence} demonstrated that many of the results of dynamical diffraction can be reproduced in the Laue case using a quantum information (QI) model, in which neutrons travel through a quantum Galton board where every peg corresponds to the application of a unitary operator on the neutron state. The intensity profiles predicted by the standard results of dynamical diffraction were reproduced with accuracy depending on the amount of layers used to model the crystal thickness. In this work we show that in the Laue case the model output reduces exactly to the form predicted by dynamic diffraction theory in the spherical incident wave case when the model parameters are taken to their appropriate limits. Additionally, we show and discuss how the model can be extended to the Bragg geometry, and that it is consistent with experimental data in complex mixed Laue-Bragg geometries where dynamic diffraction is not able to provide an analytical solution. This adaptation to new geometries is a proof of concept that this computational method is a promising approach to accurately describe complex dynamical diffraction problems.
Hence, the QI model shows promise to become indispensable for the design of novel neutron optical elements which promise to push the current limits of neutron science.

\section{Dynamical Diffraction, Laue Case}

\subsection{The Takagi-Taupin Equations}
An alternative approach to solving problems involving dynamical diffraction effects in lightly distorted crystals was developed for X-rays by Takagi and Taupin in 1962~\cite{Takagi:a03704}. In this work, we will show the equivalence between this model and the QI model. The principal results are stated in this section while a full derivation is shown in Appendix 1. The coordinate system used is shown in Fig. \ref{fig:main_diagram}.

\begin{figure*}
    \centering
    \includegraphics[width = 0.9 \textwidth]{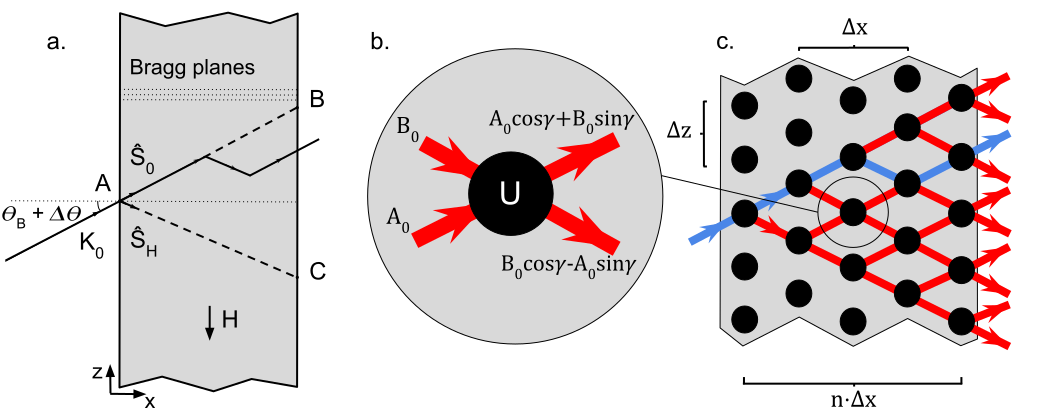}
    \caption{A side to side comparison of the Laue diffraction geometry in crystals with the mechanism behind the QI model for DD, in the Laue case. a. The real-space coordinates used in the Takagi-Taupin equations. An incident neutron with wavevector $K_0$ hits the crystal at a slight deviation from Bragg angle $\theta_B + \Delta \theta$. The coordinates $\hat{S}_0$, $\hat{S}_H$ are unit vectors in the incident and diffracted directions, respectively. b. The individual nodes act as a quantum unitary gate which splits the incident beam according to the model parameter $\gamma$. c. The diffracted amplitude at a given node is composed of the summed amplitudes of all the paths which end in the diffracted direction at that node. The widths $\Delta x$, $\Delta z$ correspond to the size of the lattice spacing in simulations. Illustrated in blue is a sample path through the lattice, which undergoes two reflections.}
    \label{fig:main_diagram}
\end{figure*}

The Takagi-Taupin equations provide an expression for the neutron wavefunction at any location inside the crystal. The neutron wavefunction is bound inside the triangle $ABC$ (the Bormann triangle), where the intensity is being shifted back and forth between the transmitted and diffracted direction. As the neutron progresses along the $x$ axis, the phase difference between the paths creates self-interference which produces a beating of the intensity, most noticeably at the center of the Bormann triangle. This beating is known as Pendellösung oscillations, and occurs with period
\begin{equation}
    \Delta_H = \frac{\pi V_{cell}\cos\theta_B}{\lambda |F_H|}
    \label{eq:Pend}
\end{equation}
where $V_{cell}$ is the volume of a crystal unit cell, $\theta_B$ is the Bragg angle, $\lambda$ is the neutron wavelength and $F_H$ is the crystal structure factor. 

We are interested in finding the position dependent intensity at the exit face of the crystal. From an experimental point of view, the position intensity can be measured directly by scanning the crystal surface with a narrow slit, and recording the intensity at every slit position $z$. 
Defining the relative transverse coordinate:
\begin{equation}
    \Gamma = \frac{z}{D\tan\theta_B}
\end{equation}
\noindent where $D$ is the crystal thickness, the intensity of the diffracted and transmitted beams at the output of the crystal are found to be, respectively
\begin{align}
    I_H(\Gamma) &= \nu^2 |A_0|^2 \mathrm{J}_0^2\left(\pi \frac{D}{\Delta_H}\sqrt{1-\Gamma^2}\right) \label{I_H_TT}\\
    I_0(\Gamma) &= \nu^2 |A_0|^2 \frac{1-\Gamma}{1+\Gamma} \mathrm{J}_1^2\left(\pi \frac{D}{\Delta_H}\sqrt{1-\Gamma^2}\right)
    \label{I_0_TT}
\end{align}
where $A_0$ is the amplitude of the incident beam, and $\mathrm{J}_0$, $\mathrm{J}_1$ are the $0^\text{th}$ and $1^\text{st}$ ordinary Bessel functions of the first kind.

\subsection{Quantum Information (QI) Model}

\begin{figure}
  \centering
  \includegraphics[width=0.45\textwidth]{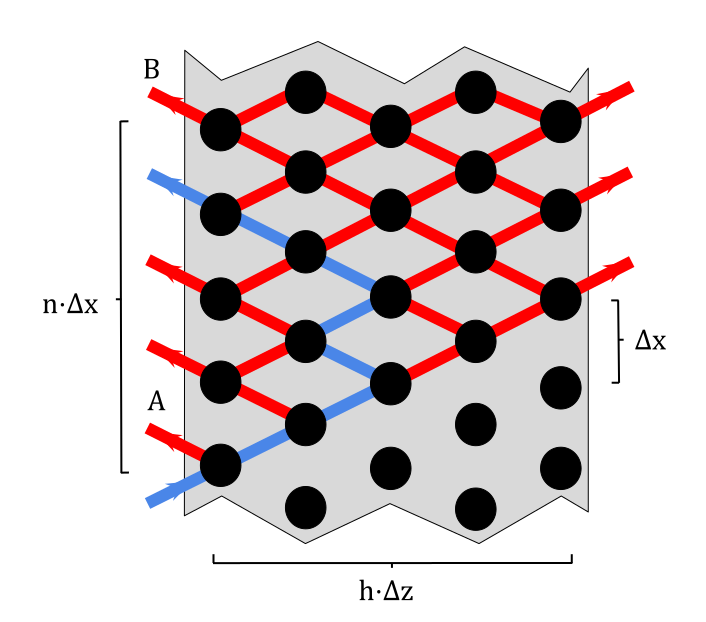}
  \caption{A lattice as it is used in the model in the Bragg case. The nodes in the lattice are functionally identical to the one presented in Fig. \ref{fig:main_diagram}b. The reflected intensity at node $n$ is a sum of the contributions from all the paths which leave the crystal from the edge at node $n$. The widths $\Delta x$, $\Delta z$ once again correspond to the lattice spacing in simulations. The height of the neutron paths through the crystal cannot exceed height $h$, which corresponds to the crystal thickness in simulation space.  Illustrated in blue is a Dyck path of length 6, bound by height 2 and containing 2 peaks. The points A and B correspond to the points of geometric reflection from the front and back face of the crystal, respectively, where the reflected intensity is typically the highest.}
  \label{DyckPaths}
\end{figure}

In the model developed in \cite{Nsofini_2016}, a perfect crystal is represented as a two-dimensional lattice of nodes, through which the incident neutron travels column by column. As shown in Fig.~\ref{fig:main_diagram}b and~\ref{fig:main_diagram}c, each node acts as a unitary operator on one part of the neutron's state, which is composed of a superposition of upwards and downwards paths at every position in the lattice. Each node corresponds to the action of one or many lattice planes upon an incident neutron, with the physical size of the node being determined by the choice of parameters. The input state to a node at position $i$ is represented by 
\begin{equation}
    \alpha_i \ket{a_i}+ \beta_i \ket{b_i}
    \quad\mathrm{or}\quad 
    \begin{pmatrix}
           \alpha_i \\
           \beta_i \\
    \end{pmatrix}
\end{equation}
where $\ket{a}$ and $\ket{b}$ are the states of the neutron going upwards (transmitted) and downwards (reflected), respectively. Evolution of the initial state is performed via the unitary time evolution operator in the interaction picture $U = e^{-i \int V_I dt / \hbar}$, where $V_I$ is the interaction potential representing the lattice. The potential integrated over the time it takes a neutron to pass through a single node is

\begin{equation}
    \langle a | V_I | b \rangle {\Delta t \over \hbar}  = \pi / 2 {\Delta x \over \Delta_H} e^{i \pmb{H} \cdot \pmb{r}} = \gamma \, i e^{i \zeta}
    \label{eqn:VIdt}
\end{equation}

\noindent
where $\Delta x = 2 m \, \Delta t / (\hbar K_x)$  is twice the distance between nodes along the Bragg planes (Figs.~\ref{fig:main_diagram}c and~\ref{DyckPaths}) with $K_x = (2 \pi / \lambda) \cos \theta_B $ the component of internal neutron wavevector also along the Bragg planes; and the phase factor encodes the global translation of the lattice. The extra factor of $i$ is inserted for convenience and corresponds to translating the lattice by one fourth of the Bragg plane spacing. Noting that $\langle a | V_I | b \rangle = \langle b | V_I | a \rangle^*$, the full time-evolution operator over one node $U_i = e^{-i \int \mathcal{H} dt / \hbar}$ is

\begin{equation}
\begin{aligned}
    U_i &= \sum_{n=0}^\infty {1 \over n !} \left ( \begin{matrix} 0 & \gamma \, e^{i \zeta} \\  - \gamma \, e^{-i \zeta} & 0 \end{matrix} \right )^n \\
    U_i &= \left ( \begin{matrix} \cos \gamma & e^{i \zeta} \sin \gamma \\
    - e^{-i \zeta } \sin \gamma & \cos \gamma
    \end{matrix}  \right )
\label{eqn:UiMatrix}
\end{aligned}
\end{equation}

\noindent
The unitary describing neutron propagation to the next layer of nodes

\begin{equation}
    U_i = \ket{a_{i+1}}(t_a \bra{a_i} + r_b \bra{b_i})+\ket{b_{i-1}}(r_a \bra{a_i}+t_b\bra{b_i})
\end{equation}

\noindent
then has coefficients

\begin{equation}
\begin{split}
    t_a = e^{i\xi}\cos{\gamma},\;\; r_b = e^{i\zeta}\sin{\gamma}\\
    r_a = -e^{-i\zeta}\sin{\gamma},\;\; t_b = e^{-i\xi}\cos{\gamma}
    \label{eq:coefficients}
\end{split}
\end{equation}

\noindent
which necessarily adhere to the required normalization conditions of a unitary matrix

\begin{equation}
    |{t_a}|^2+|{r_a}|^2 = 1,\; |{t_b}|^2+|{r_b}|^2 = 1,\; t_a \overline{r_b}+r_a\overline{t_b} = 0 .
\end{equation}

The phase $\xi$ on the diagonals is not physical and thus set to zero.  The off-diagonal phase $\zeta$ associated with a global lattice translation is important to interferometer simulations \cite{nsofini2017noise}, where a relative translation of one of the diffracting optics shifts the phase of the measured interference pattern, but it is of no consequence to the simulations presented here and also set to zero.

The input to one column containing h nodes is
\begin{equation}
   {\psi_{in}} = \begin{pmatrix}
           \alpha_i \\
           \beta_i \\
    \end{pmatrix}^{{\otimes}h}
\end{equation}
where $\alpha_i$, $\beta_i$ are the inputs in the transmitted and reflected direction to the $i^\text{th}$ node. For calculation purposes, this is written as 
\begin{equation}
    {\psi_{in}}=  \begin{pmatrix}
           \vdots \\
           \alpha_i \\
           \beta_i \\
           \alpha_{i+1} \\
           \beta_{i+1} \\
           \vdots \\
    \end{pmatrix}
\end{equation}
The column operator $U_i^{\otimes h}$ is represented as a matrix, where every node has matrix representation 
\begin{equation}
M_i = 
    \begin{pmatrix}
           t_a & r_b\\
           0 & 0 \\
           0 & 0 \\
           r_a & t_b \\
    \end{pmatrix}
    \label{eq:node}
\end{equation}
and the full column operator is written as:
\\
\begin{equation}
\newcommand*{\temp}{\multicolumn{1}{c|}{0}}
\newcommand*{\tempp}{\multicolumn{1}{c|}{r_b}}
\newcommand*{\temppp}{\multicolumn{1}{c|}{t_b}}
\newcommand*{\tenp}{\multicolumn{1}{|c}{0}}
\newcommand*{\tenpp}{\multicolumn{1}{|c}{t_a}}
\newcommand*{\tenppp}{\multicolumn{1}{|c}{r_a}}
C=\left(\begin{array}{cccccc}
\cline{2-3}
&\tenpp &\tempp &0& 0 \\
&\tenp & \temp &0 & 0 \\ \cline{4-5}
&\tenp &\temp &t_a & \tempp& \hdots \\
&\tenppp & \temppp& 0& \temp\\ \cline{2-3}
&0 & \temp& 0& \temp\\
&0 & 0&\tenppp & \temppp\\ \cline{4-5}
&&\vdots&&&\ddots
\end{array}\right)
\end{equation}

For a crystal with a thickness of N nodes, the output ${\psi_{out}}$ is equal to $C^N {\psi_{in}}$, where the odd entries of ${\psi_{out}}$ correspond to the transmitted beam at each node and the even entries to the reflected beam. The beam profiles are given by discrete functions of the node height $j$:
\begin{align}
{I_H(j)} &= |\psi_{\text{out}}(2j)|^2\\
{I_0(j)} &= |\psi_{\text{out}}(2j-1)|^2
\end{align}
\subsection{Generalization of QI model to arbitrary parameters}
It has been shown previously in \cite{Nsofini_2016} that propagating a neutron inside a lattice by exciting a single node at the entrance yielded intensity profiles consistent with dynamic diffraction theory, with accuracy for a specific choice of $\gamma$ depending on the number of layers used in the simulation. 
Here, we generalize this theory to any value of $\gamma$, and show that one has a degree of freedom when choosing a combination of $\gamma$ and the crystal thickness $n$.  Furthermore, we show that the intensity profiles generated by the model exactly reduce to the spherical wave solutions of the T-T equations, equations \ref{I_H_TT} and \ref{I_0_TT}, in the appropriate limit.

To demonstrate this, we determine analytically the intensity profiles predicted by the model at the exit face of the crystal. In Fig.~\ref{fig:main_diagram}c, in blue, a path is shown through a lattice of width $n = 2$, starting at $p = 0$ (by definition) and ending on node $p = 1$. The total neutron amplitude at $p = 1$ will be a sum of the contributions from all the paths ending on that node, and thus the problem of calculating intensity profiles can be reduced to counting lattice paths. We will start by noting that counting the number of paths of half-length $n$ ending at node $p$ is equivalent to counting the number of binary words of length $2n$ with exactly $n - p$ zeros and $n+p$ ones, where these numbers represent an up or down movement, respectively (for example, the aforementioned path corresponds to the string $0010$). Since there are $n+p$ choices for the positions of the ones, there are $2n$ choose $n+p$ 
\begin{equation}
    N(p) = {2n \choose n+p}
\end{equation}
such paths. 

However, not all paths contribute equally to the final amplitude, and a given path's weight will depend on the number of reflections which it undergoes. Instead of simply counting the paths which end on a specific node, we must additionally keep track of their number of reflections. Applying a similar logic as in the simpler case, we can derive the number of paths ending on node $p$ of length $2n$ with $k$ reflections $N(n,k,p)$:
\vspace{1em}
\begin{equation}
    N(n,k,p) = \begin{dcases}
    {n-p-1 \choose k/2-1}{n+p \choose k/2} & k \text{ even} \\
    {n-p-1 \choose (k-1)/2}{n+p \choose (k-1)/2} & k \text{ odd}
    \end{dcases}
\end{equation}
or, alternatively, for arbitrary k
\begin{align}
    N(n,2k,p) &=  {n-p-1 \choose k-1}{n+p \choose k}\\
    N(n,2k+1,p) &= {n-p-1 \choose k}{n+p \choose k}
\end{align}

Summing over all the paths ending at node $p$ and giving every path the appropriate amplitudes from equation \ref{eq:coefficients} gives us the expression for the neutron amplitude profile at the exit face of the crystal. The same paths contribute to the diffracted and transmitted intensities, up to one final reflection on the last layer. The diffracted and transmitted amplitude profile are found to be
\begin{align}
   \begin{split}
    \psi_H(p,n) =  &\sum_{k=0}^{n-|p|}(-1)^{k+1} \sin^{2k+1}\gamma\cos^{2(n-k)}\gamma \\
     &\times {n+p \choose k}{n-p \choose k}
    \label{eq:psi_H}
    \end{split}
    \\
    \begin{split}
    \psi_0(p,n) =  &\sum_{k = 0}^{n-|p|} (-1)^k \sin^{2k}\gamma \cos^{2(n-k)+1}\gamma\\
    &\times {n-p-1 \choose k-1} {n+p+1 \choose k}
     \label{eq:psi_0}
     \end{split}
\end{align}
$\psi(p)$ ranges from $p = -n$ to $p = n$. For small $n$, $\psi(p)$ has a low resolution and is a poor match to the theoretical predictions. To increase the resolution of $\psi$ while keeping the crystal thickness finite, it is necessary to ensure that the scale of the interactions decreases proportionally, so that the effective thickness of the crystal remains constant. This can be achieved by considering the limit where $\gamma\to 0$ and $n\cdot\gamma$ is kept constant, where we are able to show that the intensities $I_{0,H} = \psi_{0,H}\psi_{0,H}^*$ take the form 
\begin{align}
     I_H(p) &= {\gamma}^2\mathrm{J}^2_0(2n{\gamma}\sqrt{1-p^2/n^2})
     \label{I_H_an}\\
    I_0(p) &=  \gamma^2\frac{n+p}{n-p} \mathrm{J}^2_1(2n\gamma \sqrt{1-p^2/n^2}) \label{I_0_an}
\end{align}

Comparing Eqs.~\ref{I_H_an} and \ref{I_0_an} to Eqs.~\ref{I_H_TT} and \ref{I_0_TT} we can note that they are equivalent when we set $\Gamma = p/n$ (from its definition), $|A_0^2| = 1$ and $n\cdot \gamma =(\pi/2)\frac{D}{\Delta_H}$, as expected from equation \ref{eqn:VIdt}.

\subsection{Determining simulation parameters from experimental variables}

Since $\gamma$ and the number of simulation bi-layers $n$ are related to the crystal parameters by

\begin{equation}
    n \cdot \gamma = \pi/2\frac{D}{\Delta_H}
    \label{mainrelation}
\end{equation}
there is a degree of freedom when choosing the parameters when simulating a given experiment. One can sacrifice accuracy for speed by decreasing the number of layers $n$, as long as $\gamma$ is adjusted such that the relation in Eq.~\ref{mainrelation} is maintained. While the exactness of the model output increases as $\gamma \to 0$ and $n \to \infty$, results are already an excellent match to the theory when $\gamma$ is on the order of $\pi/100$. In this case, a crystal with a Pendellösung thickness $D/\Delta_H$ of 100 would be be composed of 5000 lattice columns, which corresponds to 10000 (10000x10000) sparse matrix multiplications which is a simple task for a modern computer. The simulation output must also be interpreted differently depending on the choice of parameters. The effective size of a simulation layer depends on the crystal thickness $D$, as well as the number of bi-layers $n$
\begin{equation}
    \Delta x \cdot n = D
\end{equation}
 and the lattice spacings in both axes are related through the Bragg angle
 \begin{equation}
     \frac{\Delta z}{\Delta x} = \tan \theta_B
 \end{equation}
 Since the simulated intensity is specified at each node, the spatial coordinate must be scaled by a factor of $\Delta z$. 
 By substituting the definition of $\Delta_H$ into Eq.\ref{mainrelation}, we obtain an expression for $\gamma$ and $\Delta x$ in terms of crystal characteristics
 
\begin{equation}
    \frac{\gamma}{\Delta x} = \frac{d |F_H|}{ V_{cell}}
    \label{theta}
\end{equation}
where $d$ is the distance between Bragg planes and $V_{cell}$ is the volume of a unit cell in the crystal. From this expression, we can observe that in the small $\gamma$ limit, variations in the value of $\gamma$ are analogous to variations in the Bragg plane distance inside the crystal, such as those resulting from strains or deformations. These effects are a computational challenge in the standard theory of dynamic diffraction, while this model offers an approach to solve these problems without the need for complex calculations. Depending on one's choices for the model parameters, the simulated profiles can often be produced very quickly, with high accuracy, and without the need for complex analytical calculations.

\section{QI model, Bragg Case}

To extend the model to the Bragg case, we introduce empty nodes, consisting of the ``transmission matrix''
\begin{equation}
T = 
    \begin{pmatrix}
           1 & 0\\
           0 & 0 \\
           0 & 0 \\
           0 & 1 \\
    \end{pmatrix}
    \label{eq:node}
\end{equation}
to create regions of the simulation environment where the neutron is propagating through empty space. It then becomes possible to simulate Bragg diffraction by filling only a segment of the simulation space with crystal nodes, and the rest with empty space in which we place a detector to keep track of the intensity being reflected from the crystal.

\begin{figure*}
    \makebox[\linewidth]{
        \includegraphics[width=\textwidth]{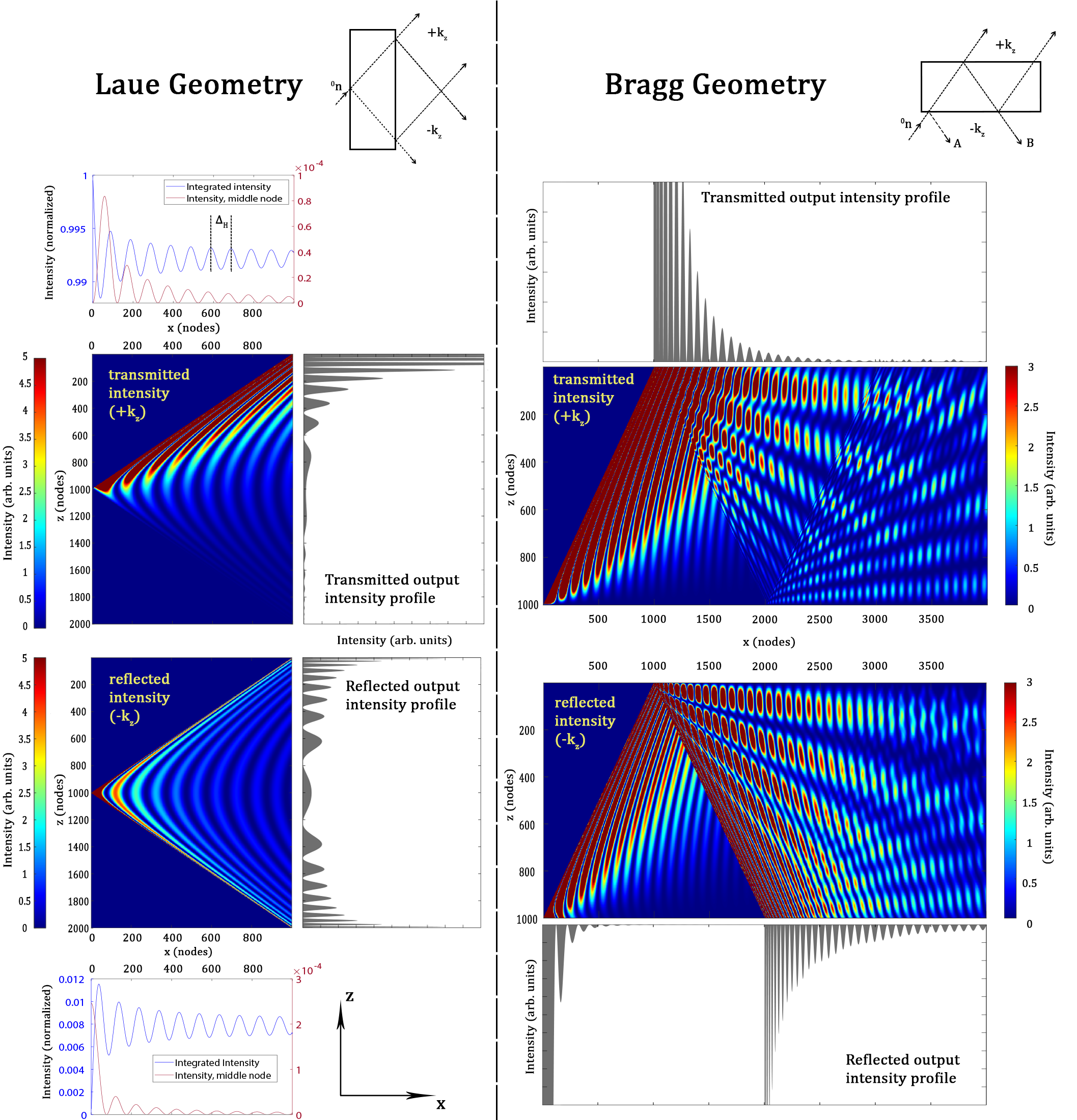}
    }
    \caption{The intensity distributions inside the crystal for the Laue case (left) and Bragg case (right). The top row figures are for the transmitted path (post selected on $+k_z$ momentum) and the bottom row figures are for the reflected path (post selected on $-k_z$ momentum). For each case we plot the output intensity profiles corresponding to the intensity at the end nodes. Lastly, the integrated intensities for the Laue case are plotted under the crystal figures showing the Pendellösung oscillations with period $\Delta_H$.}
    \label{fig:intensities}
\end{figure*}

\begin{figure}
    \centering
    \includegraphics[width = 0.48\textwidth]{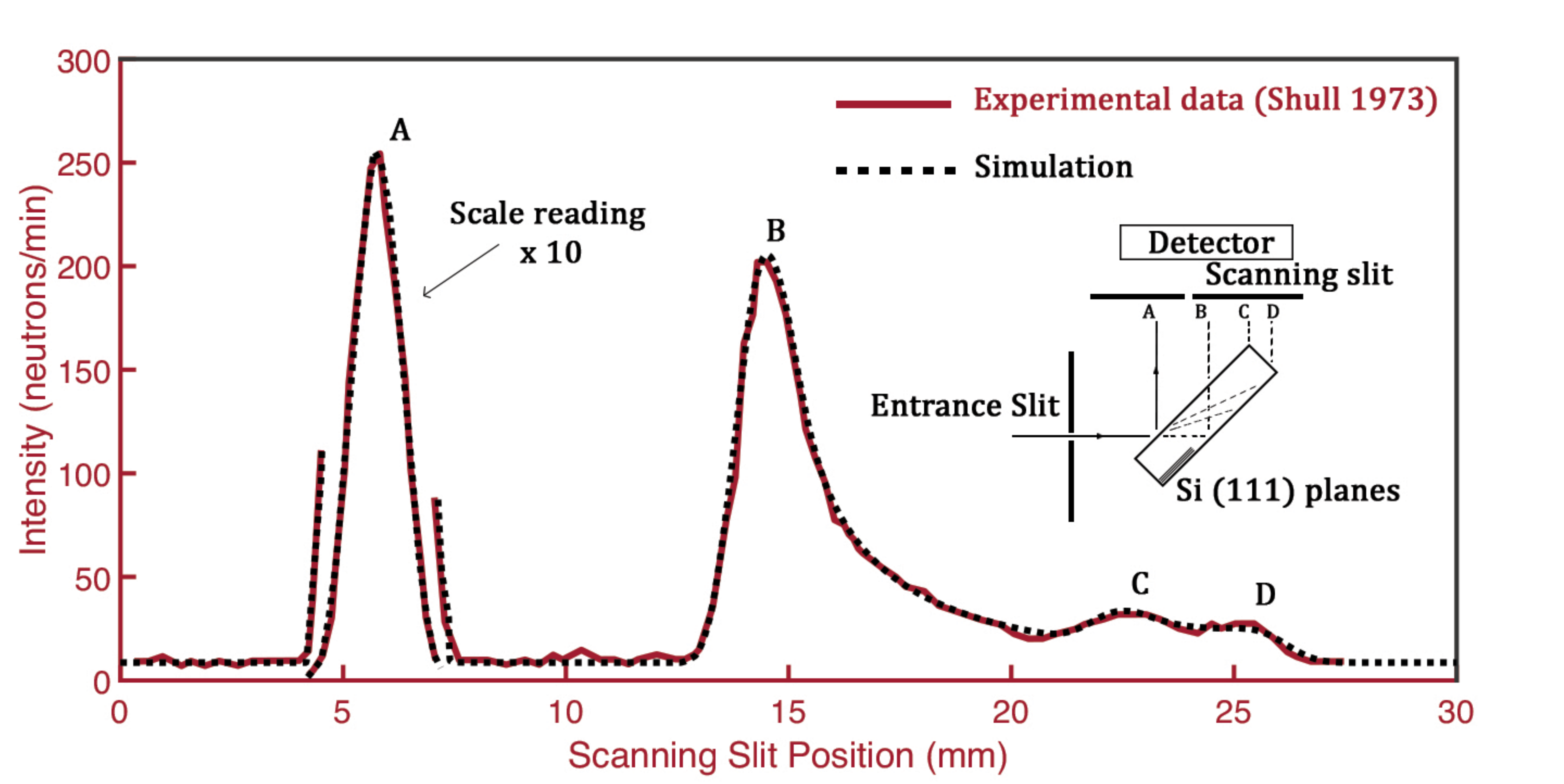}
    \caption{In red, full line: A Bragg-reflected intensity distribution measured at the exit face of a crystal by use of a scanning slit, from \cite{Shull_perfect_crystals}. In black, dotted line: A simulated intensity profile in the Bragg case, where we have chosen the parameters according to Eqn. \ref{mainrelation} and Shull's experimental parameters. The simulation output was convolved with the shape of the A peak to account for experimental effects such as slit width and beam momentum distribution. The points A, B, C and D correspond to the geometric reflection points as shown in the inset diagram.}
    \label{fig:bragg_data1}
\end{figure}
By doing so, we obtain an intensity profile for the diffracted beam. We would like to obtain an analytical expression for the reflected intensity in the Bragg case like we did for the Laue case. Since the neutron never re-enters the crystal after leaving it, we can see that this problem is equivalent to counting the number of Dyck paths \cite{CHOMSKY1959118} with some length $n$, a fixed number of peaks $k$ and a maximal height $h$. A Dyck path is a lattice walk starting at $(0,0)$ which only allows movements of $(+1,+1)$ and $(+1,-1)$, and never drops below the $x$ axis. In Fig.~\ref{DyckPaths}, we illustrate in blue that a path through a lattice in the Bragg geometry is equivalent to a Dyck path.

The total number of Dyck paths of length $2n$ is given by the Catalan numbers
\begin{equation}
    C_n = \frac{1}{n+1}{2n \choose n}.
\end{equation}
However, similar to the Laue case, the weight associated with each path depends on the number of reflections which it has undergone. Because the paths must leave from the same face through which they entered, the number of reflections is always odd and we can simply count the number of peaks $k$ of each path, defined as a local maximum in path height. The number of Dyck paths of length $2n$ with exactly $k$ peaks is given by

 \begin{equation}
        N(n,k) = \frac{1}{n}{n \choose k}{n \choose k-1}
        \label{eq:Narayana}
\end{equation}

\noindent which correspond to the Narayana numbers. If the crystal thickness was infinite, this would be enough to derive an expression for the reflected intensity everywhere. However, in the finite crystal case, starting at $n = h$, some of the fewer-peaked paths will leave the crystal through the top edge. These paths generally have a higher weight in the small $\gamma$ limit due to the factor of $\sin\gamma$ introduced on a refection, and therefore cannot be neglected. For a complete description, we require an expression for the number of Dyck paths of length $2n$, with exactly $k$ peaks and which are bound above by height $h$, which we will denote $H(n,k,h)$. Unfortunately, there is no known closed form for these numbers, but it is possible to derive a recursion relation which allows for any one of them to be computed. We divide a path from $(0,0)$ to $(2n+2,0)$ into two sections, from $0$ to $2i$ and $2i$ to $2n+2$, where $(2i,0)$ is the last point at which the path returns to the $x$ axis before it ends. There are $n$ possibilities for $i$, where $i = 0$ means the path does not return to the $x$ axis between the first and last point. There are $\sum_{j=0}^i H(i,j,h)$ such possible paths. After the path touches the $x$ axis at $x = 2i$, the next movement is necessarily upwards, and the final movement from $(2n,1)$ to $(2n+2,0)$ is necessarily downwards. Furthermore, this second path will never touch the $x$ axis again, and will never go above height $h$: we can therefore describe it as a path of half-length $n-i$ and bounded by height $h-1$. Because the number of peaks of both halves must add up to $k$, and there are $n$ choices for $i$, the total number of paths $H(n+1,k,h)$ is given by
\begin{equation}
    H(n+1,k,h+1) = \sum_{i=0}^n \sum_{j=0}^i H(i,j,h)H(n-i,k-j,h-1)
    \label{recursion}
\end{equation}
With initial conditions 
\begin{itemize}
    \item $H(0,k,h) = \delta_{k0}$
    \item $H(n,0,h) = \delta_{n0}$
    \item $H(n,k,0) = \delta_{n0}\delta_{k0}$
\end{itemize}

Using the aforementioned Narayana numbers and the same definitions as in the Laue case, we can find the reflected amplitude inside the $AB$ region (Fig. \ref{DyckPaths}) where it is unaffected by reflections off of the back face of the crystal

\begin{align}
\begin{split}
    \psi_H(n) = \sum_{k=1}^n &(-1)^{k-1} \sin^{2k-1} \gamma \cos^{2(n-k+1)} \gamma\\
    \times &\frac{1}{n}{{n}\choose{k}} {{n}\choose{k-1}}
\end{split}
\end{align}
Once again, we consider the limiting case $\gamma \to 0$ with $\gamma \cdot n$ kept constant. Here, we find the that reflected intensity is of the form 
\begin{equation}
    I_H(n) = \frac{1}{n^2}\mathrm{J}^2_1(2\gamma n)
    \label{eq:BraggIH}
\end{equation}
It has been shown experimentally that there is a secondary reflection peak on the point of geometrical reflection $n = h$, where $h$ is the crystal thickness. However, the intensity for $n < h$ is independent of $h$ since the paths have not yet had the chance to reach the top of the crystal. In this sense, equation \ref{eq:BraggIH} is a good match for experimental data when simply looking at the primary reflection peak. Furthermore, it is equivalent to the analytical solution found in \cite{abov2002dynamical} for the same region.

The similarities and differences of dynamical diffraction in the Laue case and Bragg case can be contrasted by examining the intensity inside the crystals. Fig.~\ref{fig:intensities} shows the intensities inside the crystal on the transmitted path or the reflected path. One can observe the oscillation pattern inside the crystal that leads to the Pendellösung oscillations as well as the output intensity profiles corresponding to the intensity at the last node. 

\begin{figure}
    \centering
    \includegraphics[width = 0.48\textwidth]{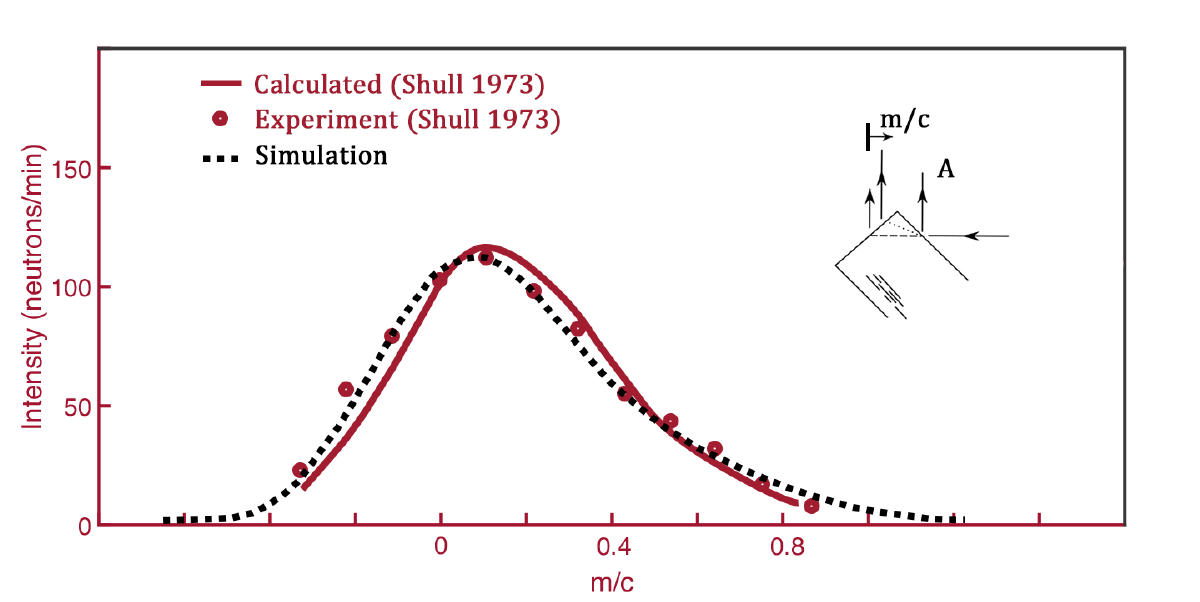}
    \caption{In red: The neutron intensity profile reflected from the corner of a Bragg crystal, from \cite{Shull_perfect_crystals}. The profile calculated by Shull is displayed as a full red line, and the corresponding experimental data points are showed. $m/c$ is a parameter describing the position of the scanning slit~\cite{Shull_perfect_crystals}. In black, dotted: The QI model simulation where the parameters were varied to obtain a good match with the experimental data. The output intensity profile was convoluted with the same profile as in Fig.~\ref{fig:bragg_data1}.}
    \label{fig:trianglefig}
\end{figure}

The diffracted neutron intensity in the Bragg case was measured by C.G. Shull and colleagues \cite{Shull_perfect_crystals} using a scanning slit to determine the beam profile exiting a crystal. In DD theory the Bragg case has 100 \% reflectivity for neutrons falling within a narrow angular range called the Darwin width 
\begin{eqnarray}
\theta_D = \frac{\lambda^2|F_H|}{\pi V_\mathrm{cell} \sin^2(2\theta_B)}
\end{eqnarray}
which is typically on the order of an arcsecond.  Neutrons outside this range propagate through the crystal and can reflect off the back face ultimately exiting from the front.  Those neutrons exiting the crystal in  this way are spatially  displaced from the primary diffraction peak by an amount $2t/\tan(\theta_B)$. In Ref.~\cite{Shull_perfect_crystals}, neutrons ($\lambda = 4.43$) were directed at a silicon crystal with $\theta_B = 44.9\degree$.  The results from \cite{Shull_perfect_crystals} are shown in Fig.~\ref{fig:bragg_data1}. 
The primary peak (labeled A) was measured to have approximately 10 times the intensity of the secondary peak (labeled B).  Two additional small peaks were observed in locations corresponding to neutrons exiting the corners of the crystal.
 
In the same figure, we have overlaid the output of our simulation model (dashed black curve). The model parameters such as $\gamma$ and the number of nodes were calculated from the experimental parameters found in Ref. \cite{Shull_perfect_crystals} using the relationship presented in equation \ref{mainrelation} and by setting $\gamma$ to $\frac{\pi}{50}$.
The model is able to accurately simulate the features observed in the experimental intensity, such as the presence of a primary peak at the first geometrical reflection point, as well smaller secondary peaks which appear where the neutron has reflected off of the back face and corners of the crystal. To obtain a proper intensity profile from the simulations, it is necessary to account for the shape of the incoming beam. This can be accomplished by convolving the simulation output with the experimentally obtained shape of the first (A) peak. The QI model enables one to easily vary the geometry of the crystal that is being analyzed. For example, to account for any slight misalignment of the Bragg planes with respect to the crystal surface, it is possible to vary the angle of the CD side of the crystal in the simulations.  By performing a least-squares fit, it is found that good agreement is obtained when the CD side of the crystal is at an angle of $91.35 \pm 0.07\degree$ relative to the AC side. The QI model can also be applied to the simulate the data from the corner of a Bragg crystal as shown in Fig.~17 of Ref.~\cite{Shull_perfect_crystals}. The same crystal geometry/parameters (including the $91.35\degree$ corner angle) and beam characteristics as that of our Fig.~\ref{fig:bragg_data1} were used. Here it was required to estimate for the physical location of the beam entrance point w.r.t. the corner of the crystal (the ``$m/c$'' parameter) as it was not specified in Ref~\cite{Shull_perfect_crystals}. We find good agreement when the beam is set to enter the crystal 6.2mm away from the corner point. The results  displayed in Fig. \ref{fig:trianglefig} once again demonstrate that the model is a good match for experimental data in the mixed Laue-Bragg case. 

\section{Conclusion}
We have shown that the quantum information model for dynamical diffraction developed by Nsofini \textit{et al.} exactly reduces to the spherical wave solutions of the Takagi-Taupin equations in the Laue case, in the limit where the density of nodes in the simulation environment approaches infinity. However, even with a relatively small number of nodes the model is in excellent agreement with the existing results of DD theory.  Furthermore, we have shown that the model can be extended successfully to incorporate the Bragg geometry, either in the pure Bragg or the mixed Laue/Bragg case, where it is a good match for existing experimental data. This demonstrates that this model is a useful tool to approach complex diffraction problems where the theory might not allow for a full description, such as when designing novel optical elements with complex shapes or accounting for strain and incoming beam phase space considerations in precision experiment.

\section{Acknowledgements}

This work was supported by the Canadian Excellence Research Chairs (CERC) program, the Natural Sciences and Engineering Research Council of Canada (NSERC) Discovery program, and the Canada  First  Research  Excellence  Fund  (CFREF). 

\section{Appendix}

\subsection{Appendix 1: The Takagi-Taupin Equations}

For this derivation, we follow ref. \cite{rauch2015neutron}.

\vspace{5mm}

Inside the crystal, the neutron wavefield can be presented in the form of a sum of plane waves (wavepacket)
\begin{equation}
    \psi(\textbf{r}) = \sum_{h} \psi_h e^{\textbf{K}_h \cdot \textbf{r}}
    \label{psi_r}
\end{equation}

where the sum runs over the reciprocal lattice vectors ${h}$, ${K_h}$ are the wave vectors ${K_h} = {K_0} + {h}$ and ${K_0}$ is the central wave vector of the incident beam. In the principal case of interest where the wavefield is composed of two strong waves in the incident and diffracted direction, equation \ref{psi_r} becomes
\begin{equation}
    \psi(\textbf{r}) = \psi_0 e^{\textbf{K}_0 \cdot \textbf{r}} + \psi_H e^{ \textbf{K}_H \cdot \textbf{r}} 
\end{equation}
Where ${H}$ is the reciprocal lattice vector normal to the Bragg planes. 
In contrast to standard dynamic diffraction theory, we now let $\psi_0$ and $\psi_H$ be slowly varying functions of position inside the crystal. $\psi({r})$ must obey the Schrödinger equation
\begin{equation}
    (\nabla^2 + k_0)^2 \psi({r}) = \nu({r}) \psi({r})
    \label{schrodinger}
\end{equation}
We define the coordinate vectors ${S}_0, {S}_H$ as shown in Fig.\ref{fig:main_diagram}
\begin{align}
    S_0 &= \frac{1}{2} \left(\frac{x}{\cos\theta_B} + \frac{z}{\sin\theta_B}\right)\\
    S_H &= \frac{1}{2} \left(\frac{x}{\cos\theta_B} - \frac{z}{\sin\theta_B}\right)
\end{align}
as the spatial coordinates parallel to the direction of ${K_0}, {K_H}$, and note that the magnitude of ${K_H}$ can be expressed in terms of the (small) misset angle of the incident wave with respect to the Bragg angle 
\begin{equation}
    K_H^2 \approx K_0^2[1-2\Delta\theta \sin(2\theta_B)]
\end{equation}

Furthermore, we define $\beta = K_0 \Delta\theta \sin (2\theta_B$) as a function of the misset angle, while $\nu_0$ and $\nu_H$ are the reduced Fourier components of the potential $\nu({r})$ . We now make an ansatz on the amplitudes $\psi_{0,H}$
\begin{align}
    \psi_0(S_0,S_H) &= e^{-i\nu_0(S_0+S_H) + i\beta S_H} U_0(S_0,S_H) \label{U0}\\
    \psi_H(S_0,S_H) &= e^{-i\nu_0(S_0+S_H) + i\beta S_H} U_H(S_0,S_H)
    \label{UH}
\end{align}
where the functions $U_{0,H}$ are simply the transmitted and diffracted amplitudes, up to a position-dependant phase. Substituting equations \ref{U0} and \ref{UH} into \ref{schrodinger} yields a pair of differential equations for the amplitudes
\begin{align}
    \frac{\partial U_0}{\partial S_0} &= -i \nu_{-H}U_H \label{eq:U_0}\\
    \frac{\partial U_H}{\partial S_H} &= -i \nu_{H}U_0 \label{eq:U_H}
\end{align}
This pair of differential equations describes the amplitude current between the two principal waves inside the crystal as they are continuously scattered back into each other. It is already somewhat intuitive that these equations are in effect the continuous case of the quantum information model. One could imagine that to solve these equations numerically, we would determine some initial conditions on $U_{0,H}$ and keep track of their value while proceeding in small increments of position. 

These equations were solved by Werner \textit{et al.} in 1986 \cite{Werner1986}. The general solution for $U_H$ is 
\begin{equation}
    U_H(S_0,S_H) = \hspace{-1mm} \sum_{n = -\infty}^{\infty} a_n \left(\frac{S_0}{S_H}\right)^{n/2}\hspace{-2mm} \mathrm{J}_n\left(2\nu\sqrt{S_0 S_H}\right)
\end{equation}
where $\mathrm{J}_n$ is the $n^{th}$ Bessel function of the first kind, $\nu^2 = \nu_H \nu_{-H}$, and the coefficients $a_n$ are determined by the initial conditions. In the case where the incident beam is confined to a very narrow slit close to the entrance edge of the crystal, the incident beam can be described by the wavefunction
\begin{equation}
    \psi_i({r}) = A_0 \delta(S_H)e^{i {K_0}\cdot {r}}
    \label{incident}
\end{equation}

where $\delta$ is the Dirac delta function. Using this function as an initial condition, the solution to equations \ref{eq:U_0} and \ref{eq:U_H} becomes
\begin{align}
    U_H(S_0,S_H) &= -i\nu_H \mathrm{J}_0\left(2\nu\sqrt{S_0 S_H}\right)\\
    U_0(S_0,S_H) &= \nu \sqrt{\frac{S_0}{S_H}} \mathrm{J}_1\left(2\nu\sqrt{S_0 S_H}\right)
\end{align}
The intensity profile of the neutron after being diffracted through a crystal was measured by Shull \cite{shull_pendellosung} by scanning the edge with a narrow slit and counting them as a function of position. To determine what one could measure with such a setup in the case of our incident beam, we must determine the intensity at $x = D$, the crystal thickness.
Rather than express the intensity as a function of $z$, it is more convenient to define the parameter
\begin{equation}
    \Gamma = \frac{z}{D\tan\theta_B}
\end{equation}
and the intensities at $x=D$ are found to be
\begin{align}
    I_H(\Gamma) &= \nu^2 |A_0|^2 \mathrm{J}_0\left(\pi \frac{D}{\Delta_H}\sqrt{1-\Gamma^2}\right)^2\\
    I_0(\Gamma) &= \nu^2 |A_0|^2 \frac{1-\Gamma}{1+\Gamma} \mathrm{J}_1\left(\pi \frac{D}{\Delta_H}\sqrt{1-\Gamma^2}\right)^2
\end{align}
where the constant $\Delta_H$ is the period of the Pendellösung interference effects inside the crystal, and can be expressed in terms of experimental variables like
\begin{equation}
    \Delta_H = \frac{\pi V_{cell}\cos\theta_B}{\lambda |F_H|}
\end{equation}

\bibliography{refs}

\end{document}